\documentclass{article}
\usepackage{ijcai23}

\usepackage{times}
\usepackage{soul}
\usepackage{url}
\usepackage[hidelinks]{hyperref}
\usepackage[utf8]{inputenc}
\usepackage[small]{caption}
\usepackage{graphicx}
\usepackage{amsmath}
\usepackage{amsthm}
\usepackage{booktabs}
\usepackage{algorithm}
\usepackage{algorithmic}
\usepackage{amsthm,amsmath,amssymb}
\usepackage{mathrsfs}
\usepackage{booktabs}
\usepackage{threeparttable}
\usepackage{multicol}
\usepackage{multirow}
\usepackage[switch]{lineno}
\usepackage{hyperref}
\usepackage{marvosym}
\usepackage{ifsym}

\title{Deep Unfolding Convolutional Dictionary Model for Multi-Contrast MRI Super-resolution and Reconstruction}

\author{
Pengcheng Lei$^1$
\and
Faming Fang$^1$\and
Guixu Zhang$^{1*}$ 
\And
Ming Xu$^{2}$\thanks{Corresponding author}
\affiliations
$^1$School of Computer Science and Technology, East China Normal University\\
$^2$Software Engineering Institute, East China Normal University\\
\emails
pengchenglei1995@163.com,
\{fmfang, gxzhang, mxu\}@cs.ecnu.edu.cn
}

\begin{document}

\maketitle
\begin{abstract}
    Magnetic resonance imaging (MRI) tasks often involve multiple contrasts. Recently, numerous deep learning-based multi-contrast MRI super-resolution (SR) and reconstruction methods have been proposed to explore the complementary information from the multi-contrast images. However, these methods either construct parameter-sharing networks or manually design fusion rules, failing to accurately model the correlations between multi-contrast images and lacking certain interpretations. In this paper, we propose a multi-contrast convolutional dictionary (MC-CDic) model under the guidance of the optimization algorithm with a well-designed data fidelity term. Specifically, we built an observation model for the multi-contrast MR images to explicitly model the multi-contrast images as common features and unique features. In this way, only the useful information in the reference image can be transferred to the target image, while the inconsistent information will be ignored. We employ the proximal gradient algorithm to optimize the model and unroll the iterative steps into a deep CDic model. Especially, the proximal operators are replaced by learnable ResNet. In addition, multi-scale dictionaries are introduced to further improve the model performance. We test our MC-CDic model on multi-contrast MRI SR and reconstruction tasks. Experimental results demonstrate the superior performance of the proposed MC-CDic model against existing SOTA methods. Code is available at \href{https://github.com/lpcccc-cv/MC-CDic}{https://github.com/lpcccc-cv/MC-CDic}.
\end{abstract}

\begin{figure}
	\centering
	\includegraphics[width=0.9\columnwidth]{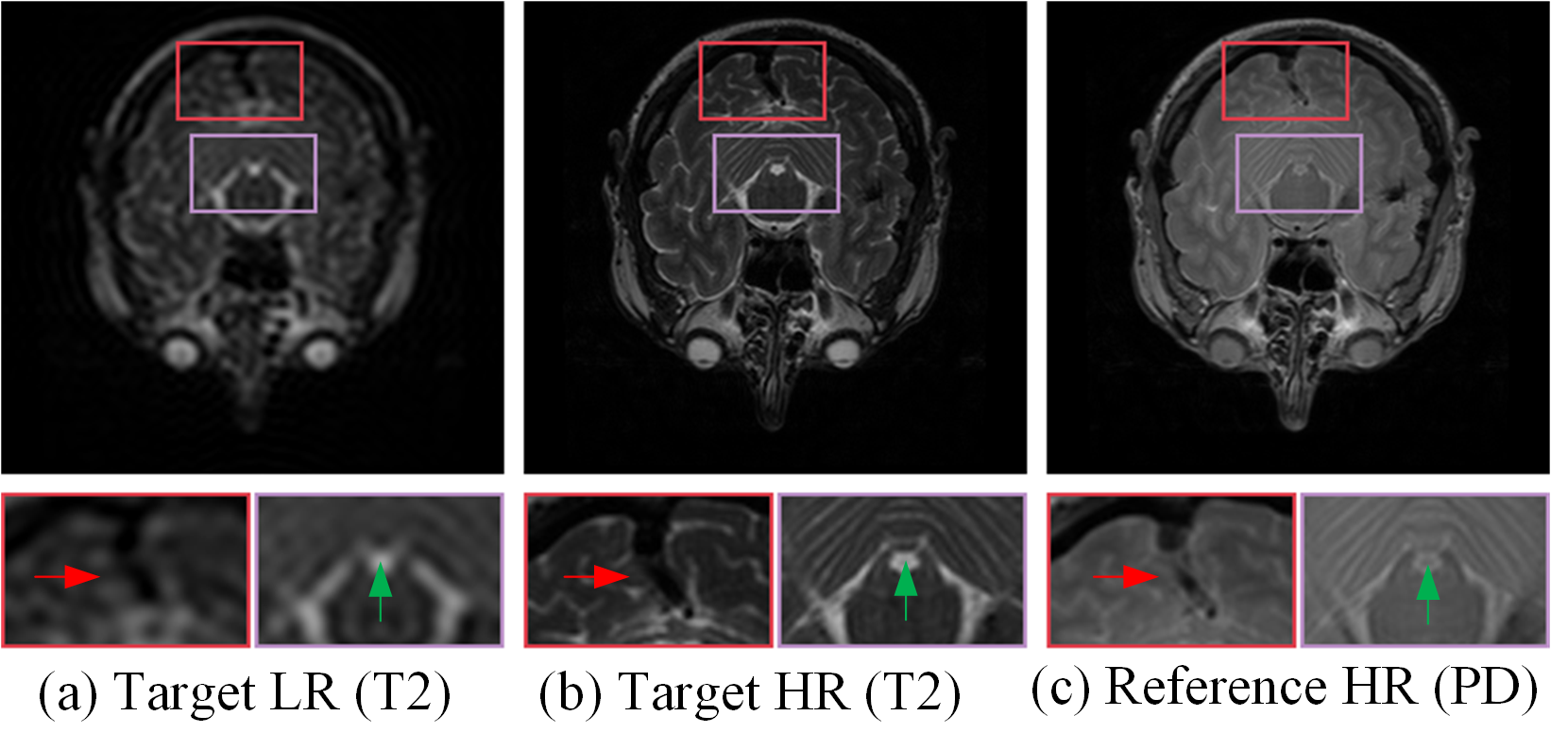} 
	\vspace{-2mm}
	\caption{The visual comparison of the multi-contrast images, i.e. T2 and PD, on IXI dataset. They share some common textures (in green), and they also have their unique contrast information and some inconsistent structures (in red).}
	\vspace{-2mm}
	\label{fig:gradient}
\end{figure}

\vspace{-2mm}
\section{Introduction}
Magnetic Resonance Imaging (MRI) is a noninvasive and non-ionizing medical imaging technique widely used for medical diagnosis, clinical analysis, and staging of disease~\cite{coupled}. However, due to the physics and physiological constraints, the MRI scanning procedure is always time-consuming. The scanning time of each patient can take up to ten minutes long, which seriously affects the patient experience and leads to high costs~\cite{CMMT}. To accelerate the MRI procedure, a commonly used approach is to undersample the k-space measurements and then use the reconstruction algorithm to restore its fully sampled imaging. 
The recent mainstream methods for accelerating MRI include MRI reconstruction and MRI super-resolution (SR). MRI reconstruction aims to remove the aliasing artifacts caused by k-space undersampling. MRI SR aims to reconstruct high-resolution (HR) MR images from low-resolution (LR) images. Both two methods can effectively improve the quality of MR images and accelerate MRI without upgrading hardware facilities. 

In the past few years, numerous MRI SR~\cite{CSNet,tcsvt,zhang2021mr} and reconstruction~\cite{wang2016accelerating,NIPSC,nitski2020cdf} methods have been proposed. However, they only use single-contrast MR images, ignoring complementary information from multi-contrast images.
In clinical practice, scanners usually produce exquisite images of body parts with different contrasts, such as T1-weighted, T2-weighted, and proton density weighted images (PD). These contrasts capture the information with the same underlying anatomy and thus exhibit similar structures~\cite{MCSR}. Furthermore, since the repetition time and echo time of different contrasts are different. For example, the scanning time of T1 and PD are significantly shorter than those of T2, so it is promising to leverage an HR (or fully sampled) reference image with a shorter acquisition time to reconstruct the modality with a longer scanning time from an LR (or undersampled) image. We name them guided-SR and guided-reconstruction.


Recently, deep convolutional neural network (DCNN)~\cite{liu2021efficient} has become a mainstream approach for solving MRI guided-SR problems.
\cite{MCSR} proposed to fuse the multi-contrast information in the high-level feature space. 
\cite{MINet} used a multi-stage integration network to explore the rich relations among different contrasts. \cite{MCMRSR,Fangmm22} designed a transformer-based framework to fully explore the long-range dependencies in both reference and target images. 
For solving the multi-contrast reconstruction problem, \cite{xiang2018deep} fused multi-contrast MR acquisitions through deep networks to speed up the reconstruction of a certain target image. \cite{sun2019deep} proposed a deep information-sharing network to jointly reconstruct multi-contrast MRI images. \cite{dar2020prior} employed an auxiliary contrast as prior information into the generator of a generative adversarial network (GAN) to accelerate MR imaging.
\cite{zhang2021temporal} applied the temporal feature fusion block based on ADMM to achieve multi-modal MRI reconstruction. 
\cite{CMMT} proposed a multi-modal transformer (MTrans) to transfer multi-scale features from the target modality to the auxiliary one.

Although these methods have gotten promising results, their network architectures are black-boxes, making it difficult to understand the underlying mechanism and lack certain physical interpretations. As is well known, the model-based method~\cite{song2019coupled} is explainable and easy to incorporate domain knowledge. Most recently, numerous well-designed model-driven DCNN methods have been proposed to solve various image restoration problems, including deraining~\cite{Wang_2020_CVPR}, debluring~\cite{liu2019deep}, image fusion~\cite{liu1} and so on. For MR images, \cite{yang2018admm,aggarwal2018modl,VNnet} proposed model-based reconstruction methods based on the compressed sensing theory. However, they are designed for single-contrast MRI reconstruction.

In this paper, we propose a novel multi-contrast convolutional dictionary (MC-CDic) model to explicitly model the correlations of multi-contrast images for guided-SR and reconstruction tasks. MC-CDic model is designed by the following two observations. First, the multi-contrast images are sampled under the same anatomical structure, they thus share some common features, e.g., edges and texture information. Second, since the two images are generated by different scan settings, they also have their unique contrast information and some unaligned inconsistent structures, as shown in Figure~\ref{fig:gradient}. \emph{To better reconstruct the target image, we hope to transfer the common texture information from the reference image to the target image while avoiding the interference of inconsistent information}. Taking these into consideration, we first extract the common features and unique features from the multi-contrast images and then fuse these features to generate the desired high-quality images. 

To extract the features from both images, firstly, we adopt the convolutional sparse coding (CSC) technique to model the multi-contrast MR images. Specifically, the multi-contrast images are decomposed into two types of features (i.e., common features and unique features) using two types of dictionaries (i.e., common filters and unique filters). 
Secondly, we employ implicit regularizations on the decomposed features and use the proximal gradient algorithm to optimize them. Thirdly, we unfold the iterative optimization process into a deep network. Especially, the unfolding process strictly follows the rules of the solution and the proximal operators are replaced by learnable residual blocks. Thus all the parameters can be learned in an end-to-end manner. Fourthly, we further employ multi-scale dictionaries to extract the multi-scale sparse representations in each iteration. 
In this way, the proposed network is not only interpretable but also powerful. Our contributions can be summarized as follows.
\begin{itemize}
	\item We propose a deep unfolding network based on convolutional dictionary learning for multi-contrast MRI SR and reconstruction. Different from existing methods that design unsophisticated fusion rules, we develop priors and constraints plugged into the model according to the characteristics of multi-contrast MR images.
	\item We build an observation model for multi-contrast MRI using the CSC technique and use the proximal gradient algorithm to solve the model. Based on the iterative solutions, we unroll the iteration steps into a multi-scale convolutional dictionary network and the proximal operators are replaced by learnable deep residual networks.
	\item The proposed model combines the good interpretability of the dictionary model with powerful deep neural networks. We test it on guided MRI SR and reconstruction tasks. Extensive experiments on the benchmark datasets demonstrate the effectiveness of the proposed model.	
\end{itemize}

\section{MC-CDic Model for MR Images}
\subsection{Model Formulation}
Let $\hat{x}_{1}\in \mathbb{R}^{m\times n}$ and $\hat{x}_{2}\in \mathbb{R}^{m\times n}$ denote two classes of HR (or fully-sampled) MR images with different contrasts (e.g. PD and T2). The corresponding LR (or undersampled) images of $\hat{x}_{2}$ is ${x}_{2}$. For the guided-SR task, our MC-CDic model aims to restore the HR image $\hat{x}_{2}$ from LR image ${x}_{2}$ with the aid of the reference HR image $\hat{x}_{1}$. Since the two contrast images are sampled under the same anatomical structure, they share some common features. Since they are sampled under different scan setting, they also have their unique features. Based on these observations, inspired by~\cite{CUNet}, we model the multi-contrast images as:
\begin{equation}
	\label{model_x1}
	\hat{x}_{1} = \sum_{k}d_k^c \otimes c_k + \sum_{k}d_k^u \otimes u_k,
\end{equation}
\vspace{-2mm}
\begin{equation}
	\label{model_x1}
	{x}_{2} = \sum_{k}h_k^c \otimes c_k + \sum_{k}h_k^v \otimes v_k,
\end{equation}
where $\otimes$ denotes the convolutional operation. $\{d_k^c\}_{k=1}^K$ and $\{h_k^c\}_{k=1}^K$ are the common filters of $\hat{x}_1$ and $x_2$, $\{d_k^u\}_{k=1}^K$ and $\{h_k^v\}_{k=1}^K$ are the unique filters of $\hat{x}_1$ and $x_2$, $\{c_k\}_{k=1}^K$ represent the common features between $\hat{x}_1$ and $x_2$, $\{u_k\}_{k=1}^K$ and $\{v_k\}_{k=1}^K$ represent the unique features of $\hat{x}_1$ and $x_2$. $K$ denotes the channel numbers of the feature representations. Based on the decomposed common features and unique features, the reconstructed HR image can be represented as:
\begin{equation}
	\label{rec_x}
	x_{2}^{'} = \sum_{k}q_k^c \otimes c_k + \sum_{k}q_k^v \otimes v_k,
\end{equation}
where $\{q_k^c\}_{k=1}^K$ and $\{q_k^v\}_{k=1}^K$ are the common filters and the unique filters for reconstructing the HR images $x_{2}^{'}$.

For the above modeling, the main problem is to extract the common features and the unique features from the input images. Assuming all the filters are known, they can be solved by optimizing the following problem:
\begin{small}
\begin{equation}
	\label{obj_func}
	\begin{aligned}
		\underset{\{C,U,V\}}{\operatorname{min}}
		&\frac{1}{2}\|\hat{x}_{1}-\sum_k({d}_k^c \otimes {c}_k+{d}_k^u \otimes {u}_k)\|_F^2   \\
		&+\frac{1}{2}\|{x}_{2}-\sum_k({h}_k^c \otimes {c}_k+{h}_k^v \otimes {v}_k)\|_F^2  \\
		&+\lambda_c\psi_c(C) + \lambda_u\psi_u(U) + \lambda_v\psi_v(V),
	\end{aligned}
\end{equation}
\end{small}
where $\{C,U,V\}\in \mathbb{R}^{m\times n\times K}$ are the stacked features response to $\{c,u,v\}_{k=1}^K$, $\lambda_c$, $\lambda_u$, $\lambda_v$ are the trade-off parameters, $\psi_c(\cdot)$, $\psi_u(\cdot)$, $\psi_v(\cdot)$ are the regularizers to deliver the prior information to the feature maps.

\begin{figure*}[t]
	\centering
	\includegraphics[width=2\columnwidth]{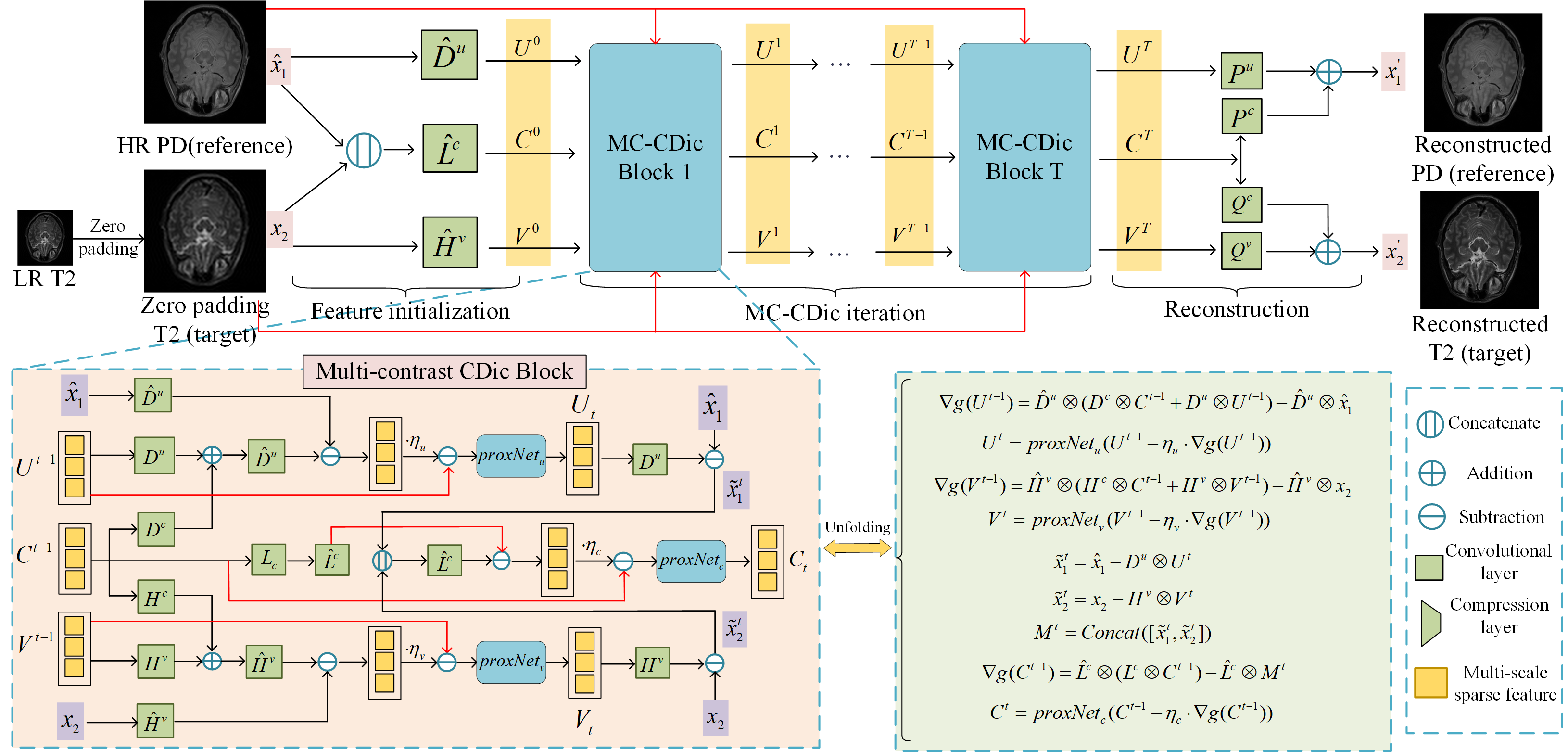} 
	\caption{The overall structure of the proposed multi-contrast convolutional dictionary (MC-CDic) model for the guided-SR task.}
	\label{main_flow}
	\vspace{-2mm}
\end{figure*}

\subsection{Optimization Algorithm} 
To solve this problem, we alternately update each variable with other variables fixed. It leads to the following three subproblems:
\begin{small}
\begin{equation}
	\label{sub-u}
	\underset{U}{\operatorname{min}} \frac{1}{2}\|\hat{x}_{1}-\sum_k({d}_k^c \otimes {c}_k+{d}_k^u \otimes {u}_k)\|_F^2 +\lambda_u\psi_u(U),
\end{equation}
\end{small}
\vspace{-2mm}
\begin{small}
\begin{equation}
	\label{sub_v}
	\underset{V}{\operatorname{min}} \frac{1}{2}\|{x}_{2}-\sum_k({h}_k^c \otimes {c}_k+{h}_k^v \otimes {v}_k)\|_F^2 + \lambda_v\psi_v(V),
\end{equation}
\end{small}
\vspace{-2mm}
\begin{small}
\begin{equation}
	\begin{aligned}
		\label{sub_c}
		&\underset{C}{\operatorname{min}} \frac{1}{2}\|\hat{x}_{1}-\sum_k({d}_k^c \otimes {c}_k+{d}_k^u \otimes {u}_k)\|_F^2 \\
		&+\frac{1}{2}\|{x}_{2}-\sum_k({h}_k^c \otimes {c}_k+{h}_k^v \otimes {v}_k)\|_F^2 +\lambda_c\psi_c(C).
	\end{aligned}
\end{equation}
\end{small}
Next, we optimize these three subproblems separately using the proximal gradient method. The details are as follows:

\paragraph{Updating $U$.} The unique feature of $\hat{x}_{1}$, $U$ can be updated by solving the quadratic approximation~\cite{beck2009fast} of the subproblem (\ref{sub-u}) as:
\begin{small}
\begin{equation}
	\label{proximal_sub_u}
	\underset{U}{\operatorname{min}} \frac{1}{2}\|U-(U^{t-1}-\eta_u\triangledown f(U^{t-1}))\|_F^2+\eta_u\lambda_u\psi_u(U),
\end{equation}
\end{small}
where $U^{t-1}$ is the updating result of the last iteration, $\eta_u$ is the stepsize parameter, and
\begin{small}
\begin{equation}
	f(U^{t-1})=\frac{1}{2}\|\hat{x}_{1}-\sum_k({d}_k^c \otimes {c}_k^{t-1}+{d}_k^u \otimes {u}_k^{t-1})\|_F^2,
\end{equation}
\end{small}
\vspace{-2mm}
\begin{equation}
	\label{grad_u}
	\triangledown f(U^{t-1}) = D^u\otimes^{\top}(\sum_k({d}_k^c \otimes {c}_k^{t-1}+{d}_k^u \otimes {u}_k^{t-1})-\hat{x}_{1}),
\end{equation}
where $\otimes^{\top}$ denotes the transposed convolution, $D^u\in \mathbb{R}^{1\times n\times n\times K}$ is a 4-D tensor stacked by $\{d_k^u\}_{k=1}^K$. Eq.~(\ref{grad_u}) can be re-written as:
\begin{small}
\begin{equation}
	\label{grad_grad_u}
	\triangledown f(U^{t-1}) = D^u\otimes^{\top}(D^c \otimes C^{t-1}+D^u \otimes U^{t-1})-D^u\otimes^{\top} \hat{x}_{1},
\end{equation}
\end{small}
where $D^c\in \mathbb{R}^{1\times n\times n\times K}$ is a 4-D tensor stacked by $\{d_k^c\}_{k=1}^K$.
Corresponding to general regularization term~\cite{donoho1995noising}, the solution of Eq.~(\ref{proximal_sub_u}) is:
\begin{small}
\begin{equation}
	\label{solution_u}
	U^t = 
	\operatorname{prox}_{\eta_u\lambda_u}(U^{t-1}-\eta_u \triangledown f(U^{t-1})),
\end{equation}
\end{small}
where $\operatorname{prox}_{\eta_u\lambda_u}(\cdot)$ is the proximal operator dependent on the regularization term $\psi_u(\cdot)$ with respect to $U$. Instead of choosing a fixed regularizer in the model, the form of the proximal operator can be automatically learned from training data.

\paragraph{Updating $V$.} Similarly, the quadratic approximation of
the problem (\ref{sub_v}) with respect to $V$ is:
\begin{small}
\begin{equation}
	\label{proximal_sub_v}
	\underset{V}{\operatorname{min}} \frac{1}{2}\|V-(V^{t-1}-\eta_v\triangledown f(V^{t-1}))\|_F^2+\eta_v\lambda_v\psi_v(V)
\end{equation}
\end{small}
where
\begin{small}
\begin{equation}
	f(V^{t-1})=\frac{1}{2}\|{x}_{2}-\sum_k({h}_k^c \otimes {c}_k^{t-1}+{h}_k^v \otimes {v}_k^{t-1})\|_F^2,
\end{equation}
\end{small}
\vspace{-2mm}
\begin{small}
\begin{equation}
	\label{grad_v}
	\triangledown f(V^{t-1}) = H^v\otimes^{\top}((\sum_k({h}_k^c \otimes {c}_k^{t-1}+{h}_k^v \otimes {v}_k^{t-1}))-{x}_{2}).
\end{equation}
\end{small}
$H^v\in \mathbb{R}^{1\times n\times n\times K}$ is a 4-D tensor stacked by $\{h_k^v\}_{k=1}^K$. Eq.~\ref{grad_v} can be re-written as:
\begin{small}
\begin{equation}
	\label{grad_grad_v}
	\triangledown f(V^{t-1}) = H^v\otimes^{\top}(H^c \otimes C^{t-1}+\textsc{H}^v \otimes V^{t-1})-H^v\otimes^{\top} {x}_{2},
\end{equation}
\end{small}
where $H^c\in \mathbb{R}^{1\times n\times n\times K}$ is a 4-D tensor stacked by $\{h_k^c\}_{k=1}^K$. The solution of Eq.~(\ref{proximal_sub_v}) can be represented as:
\begin{small}
\begin{equation}
	\label{solution_v}
	V^t = 
	\operatorname{prox}_{\eta_v\lambda_v}(V^{t-1}-\eta_v \triangledown f(V^{t-1})),
\end{equation}
\end{small}
where $\operatorname{prox}_{\eta_v\lambda_v}(\cdot)$ is the proximal operator correlated to the regularization term $\psi_v(\cdot)$ with respect to $V$.

\paragraph{Updating $C$.} We first denote $\tilde{x}_{1}=\hat{x}_{1}-\sum_k{d}_k^u \otimes {u}_k$ and $\tilde{x}_{2}={x}_{2}-\sum_k{h}_k^v \otimes {v}_k$, Eq.~(\ref{sub_c}) can be re-written as follows:
\begin{small}
\begin{equation}
	\label{sub_sub_c}
		\underset{C}{\operatorname{min}} \frac{1}{2}\|\tilde{x}_{1}-\sum_k{d}_k^c \otimes {c}_k\|_F^2 +\frac{1}{2}\|\tilde{x}_{2}-\sum_k{h}_k^c \otimes {c}_k\|_F^2 +\lambda_c\psi_c(C)
\end{equation}
\end{small}
Then the first two terms in Eq. (\ref{sub_sub_c}) can be further combined and this yields the following optimization problem:
\begin{small}
\begin{equation}
	\label{sub_sub_sub_c}
	\underset{C}{\operatorname{min}} \frac{1}{2}\|M-\sum_k{l}_k^c \otimes {c}_k\|_F^2 +\lambda_c\psi_c(C)
\end{equation}
\end{small}
where $M$ is the concatenate of $\tilde{x}_{1}$ and $\tilde{x}_{2}$, $l_k^c$ is the concatenation of $d^c_k$ and $h^c_k$. Thus, Eq.~(\ref{sub_sub_sub_c}) can be solved by the same way as $U$ and $V$. Specifically, the quadratic approximation of this problem to $C$ is:
\begin{small}
\begin{equation}
	\label{proximal_sub_c}
	\underset{C}{\operatorname{min}} \frac{1}{2}\|C-(C^{t-1}-\eta_c\triangledown f(C^{t-1}))\|_F^2+\eta_c\lambda_c\psi_c(C)
\end{equation}
\end{small}
where
\begin{small}
\begin{equation}
	f(C^{t-1})=\frac{1}{2}\|M-\sum_k{l}_k^c \otimes {c}_k^{t-1}\|_F^2,
\end{equation}
\end{small}
\vspace{-2mm}
\begin{small}
\begin{equation}
	\label{grad_c}
	\triangledown f(C^{t-1}) = L^c\otimes^{\top}\sum_k({l}_k^c \otimes {c}_k^{t-1}-M),
\end{equation}
\end{small}
where $L^c\in \mathbb{R}^{n\times n\times K \times 2}$ is a 4-D tensor stacked by $\{l_k^c\}_{k=1}^K$. Eq.~(\ref{grad_c}) can be re-writed as:
\begin{small}
\begin{equation}
	\label{grad_grad_c}
	\triangledown f(C^{t-1}) = L^c\otimes^{\top}(L^c \otimes C^{t-1})-L^c\otimes^{\top} M,
\end{equation}
\end{small}
The solution of Eq.(\ref{proximal_sub_c}) can be represented as:
\begin{small}
\begin{equation}
	\label{solution_c}
	C^t = 
	\operatorname{prox}_{\eta_c\lambda_c}(C^{t-1}-\eta_c \triangledown f(C^{t-1})),
\end{equation}
\end{small}
where $\operatorname{prox}_{\eta_c\lambda_c}(\cdot)$ is the proximal operator correlated to the regularization term $\psi_c(\cdot)$ with respect to $C$.

\section{Deep Unfolding MC-CDic Model}
Before unfolding the above model, we first formulate some relaxations on the dictionaries. Following earlier works~\cite{simon2019rethinking,lecouat2020fully,lecouat2020flexible,MUSC}, we untie the convolutional dictionaries from their transposed convolutional dictionaries in Eq.~(\ref{grad_grad_u}), (\ref{grad_grad_v}) and (\ref{grad_grad_c}) for better performance. To be specific, we initialize the convolution dictionary $D^u\otimes$ identically to $D^{u}\otimes^{\top}$ but are allowed to evolve independently during training. 
For convenience, we use $\widehat{D}^{u}\otimes$ to replace $D^{u}\otimes^{\top}$.
Similarly, $H^{v}\otimes^{\top}$ and $L^{c}\otimes^{\top}$ are also replaced by $\widehat{H}^{v}\otimes$ and $\widehat{L}^{c}\otimes$ in our deep unfloding model.

\begin{figure*}[t]
	\centering
	\includegraphics[width=1.9\columnwidth]{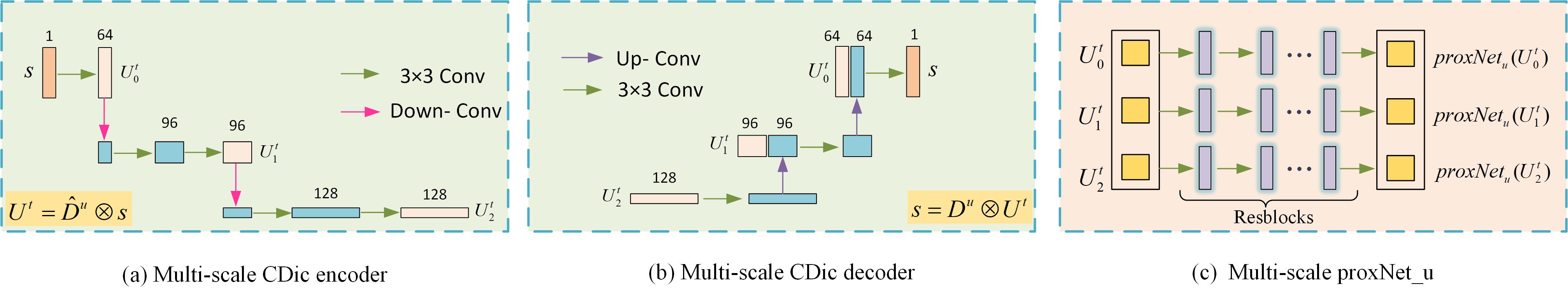} 
	\caption{The structure of the proposed multi-scale dictionaries (for updating $U^t$). (a) multi-scale CDic encoder, (b) multi-scale CDic decoder, and (c) multi-scale proximal network. $U_t=\{U_0^{t},U_1^{t},U_2^{t}\}$ contains a list of multi-scale representations.}
	\label{fig:MS_CDic}
	\vspace{-1mm}
\end{figure*}

\subsection{Network Design}
\paragraph{(1) Feature Initialization Module.} We use three convolutional dictionaries to initialize the common features and the unique features from the multi-contrast images $\hat{x}_1$ and $x_2$. Specifically, we first use the dictionaries $\widehat{D}^{u}$ and $\widehat{H}^{v}$ to initialize the unique feature $U^0$ and $V^0$ respectively. Then we concatenate $\hat{x}_1$ and $x_2$, and use $\widehat{L}^{c}$ to initialize the common feature $C^0$.
\paragraph{(2) MC-CDic Iteration Module.} As shown in the figure, our model consists $T$ MC-CDic blocks, representing $T$ iterations of the algorithm for solving Eq.~(\ref{obj_func}). Specifically, we take the input images $\hat{x}_1$ and $x_2$, and the previous outputs $U^t$, $V^t$ and $C^t$ as inputs, and outputs the updated $U^{t+1}$, $V^{t+1}$ and $C^{t+1}$ in each iteration block. The iteration blocks are unfolded following the updating rules of Eq.~(\ref{solution_u}), (\ref{solution_v}) and (\ref{solution_c}). To be specific, the convolutional dictionary operations can be easily implemented by the commonly used convolutional layers in normal networks. The key issue of unrolling the algorithm is how to represent the three proximal operators $\operatorname{prox}_{\eta_u\lambda_u}(\cdot)$, $\operatorname{prox}_{\eta_v\lambda_v}(\cdot)$ and $\operatorname{prox}_{\eta_c\lambda_c}(\cdot)$. Inspired by recent unfolding networks~\cite{Wang_2020_CVPR,wang2022ada}, we employ the deep residual network (ResNet) to replace the proximal operators. Note that the $T$ MC-CDic iterations share the same parameters in our model.

Thus, at the $t^{th}$ stage, the network for updating $U$ can be built as:
\begin{small}
\begin{equation}
	\label{unet}
	\left\{\begin{array}{l}
		\triangledown f(U^{t-1}) = \widehat{D}^u\otimes(D^c \otimes C^{t-1}+ D^u \otimes U^{t-1})-\widehat{D}^u\otimes \hat{x}_{1}, \\
		U^t = 
		\operatorname{proxNet}_{u}(U^{t-1}-\eta_u \triangledown f(U^{t-1})),
	\end{array}\right.
\end{equation}
\end{small}
where $\operatorname{proxNet}_{u}(\cdot)$ denotes the ResNet with several resblocks. Similarly, the network for updating $V$ at the $t^{th}$ stage can be built as:
\begin{small}
\begin{equation}
	\label{vnet}
	\left\{\begin{array}{l}
		\triangledown f(V^{t-1}) = \widehat{H}^v\otimes(H^c \otimes C^{t-1}+H^v \otimes V^{t-1}) -\widehat{H}^v\otimes x_{2}, \\
		V^t = 
		\operatorname{proxNet}_{v}(V^{t-1}-\eta_v \triangledown f(V^{t-1})),
	\end{array}\right.
\end{equation}
\end{small}
where $\operatorname{proxNet}_{v}(\cdot)$ denotes the ResNet with several resblocks.

Based on the updated unique features $U^t$ and $V^t$, the network for updating $C$ at the $t^{th}$ stage can be built as:
\begin{small}
\begin{equation}
	\label{cnet}
	\left\{\begin{array}{l}
		\bar{x}_1^t=x_1-D^u \otimes U^t, \\
		\bar{x}_2^t=x_2-H^v \otimes V^t,\\
		M^t = Concat(\left[\bar{x}_1^t, \bar{x}_2^t\right]), \\
		\triangledown f(C^{t-1}) = \widehat{L}^c\otimes(L^c \otimes {C}^{t-1})-\widehat{L}^c\otimes M^t, \\
		C^t = 
		\operatorname{proxNet}_{c}(C^{t-1}-\eta_c \triangledown f(C^{t-1})),
	\end{array}\right.
\end{equation}
\end{small}
where $\operatorname{proxNet}_{c}(\cdot)$ denotes the ResNet with several resblocks, $M^t$ is the concatenated features of $\bar{x}_1^t$ and $\bar{x}_2^t$.
According to the updating rules of $U$, $V$ and $C$ in Eq.~(\ref{unet}), (\ref{vnet}) and (\ref{cnet}), we design the multi-contrast convolutional dictionary (MC-CDic) block. Its overall structure is shown in Figure~\ref{main_flow}. All the parameters involved can be automatically learned from training data in an end-to-end manner. 
\paragraph{(3) Reconstruction Module.}
After $T$ MC-CDic iterations, we get the final common features $C^T$ and the final unique feature $U^T$ and $V^T$. 
Then the reconstructed HR image $x_2^{'}$ can be obtained by:
\begin{equation}
		x_2^{'} = Q^c \otimes C^T + Q^v \otimes V^T,
\end{equation}
where $Q^c=\{q_k^c\}_{k=1}^K$ and $Q^u=\{q_k^u\}_{k=1}^K$ are the common filters and unique filters for reconstructing the HR images.\\
\vspace{-3mm}
\subsection{Multi-Scale Dictionaries}
The importance of multi-scale representations has been proven in numerous computer vision tasks. However, most existing CDic models only use the single-scale dictionary to extract single-scale representations, which limits their performance to a large extent. Inspired by~\cite{MUSC}, we employ the multi-scale dictionaries in our MC-CDic model to perceive the multi-scale representations. To be specific, we employ a stripped-down version of the image synthesis process of U-Net by removing all non-linearities, batch normalization, and additive biases. We regard the decoder part of the UNet as a normal CDic layer and regard the encoder part of the UNet as its transposed CDic layer. As shown in Figure~\ref{fig:MS_CDic} (a) and (b), we present the operations of $U^t=\hat{D}^{u}\otimes s$ and $s=D^u\otimes U^t$. 
The proximal network for multi-scale dictionaries is shown in Figure~\ref{fig:MS_CDic} (c). 

\subsection{Loss Function}
We employ the simple L1 loss to supervise the multi-contrast images simultaneously. It can be formulated as:
\begin{equation}
	\mathcal{L} = \alpha||x_1^{'}-\hat{x}_1||_1 + ||x_2^{'}-\hat{x}_2||_1,
\end{equation}
where $\alpha$ is a trade-off parameter to balance the importance of different contrasts. In this paper, we set $\alpha=0.1$. Note that the loss on the reference image is to ensure that the reference image can be efficiently decomposed.

\begin{figure*}[t]
	\centering
	\includegraphics[width=1.8\columnwidth]{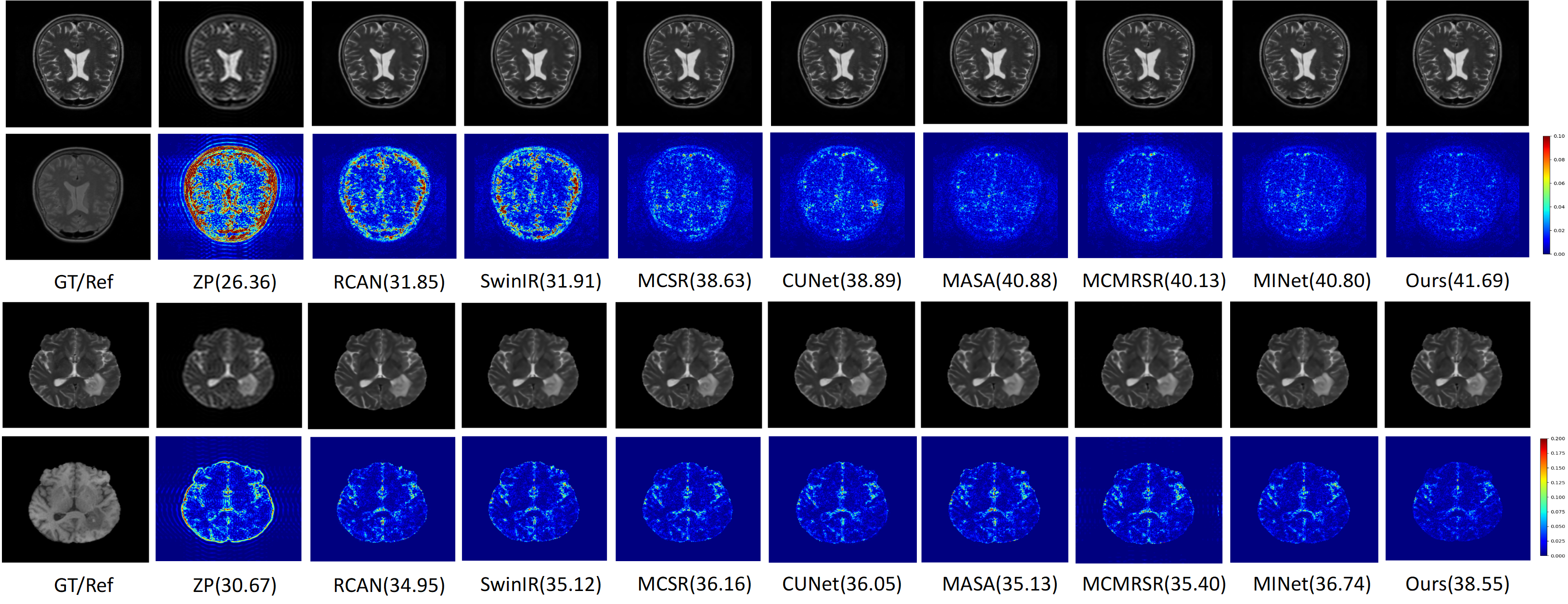} 
	\caption{Error maps of different single-contrast SR (the first three methods) and guided-SR (the next six methods) with the scale factor of $\times4$ on the IXI (PD guides T2) and BrainTS (T1 guides T2) testing sets.}
	\label{fig:ref_sr}
\end{figure*}

\section{Experiments}
\subsection{Datasets}
Following~\cite{MCSR,Fangmm22}, we employ two publicly available multi-modal MR image datasets, IXI and BraTS2018~\cite{BRATS} to validate the effectiveness of our model. IXI dataset contains 576 multi-contrast MR volumes and we employ the PD as the reference modal to guide the reconstruction of the target T2 modal. The BraTS2018 dataset contains 285 multi-contrast MR volumes and we employ the T1 as the reference modality to guide the reconstruction of the target T2 modal. We split the two datasets patient-wisely into a ratio of 7:1:2 for training/validation/testing. For each volume, we only select the middle 100 slices for our experiments. We use the cartesian sampling mask and the k-space center cropping~\cite{MCSR} mask to get the undersampled and LR images, respectively. The acceleration ratio for reconstruction and the scale factor for SR are set at 4.
\subsection{Implementation Details}
\paragraph{Training Details.} Our proposed MC-CDic model is implemented in PyTorch with an NVIDIA RTX3090 GPU. We adopt the Adam optimizer with a batch size of 6. The learning rate is set to $1\times10^{-4}$ and the models are trained 50 epochs.
We employ the peak-to-noise-ratio (PSNR), structural similarity index (SSIM), and root mean squared error (RMSE) to evaluate the model performance. The higher the PSNR and SSIM, the lower the RMSE represents the better result. \\
\paragraph{Model Details.} In our large model MC-CDic-L, the iteration stage $T$ is set to 4. The levels of the multi-scale dictionaries are set to $3$. From the first level to the third level, the channel numbers of the features are set to 64, 96, and 128. The numbers of the resblocks for the multi-scale proximal network are set to 5, 3, and 1. In our small model MC-CDic-S, we use the single-scale dictionary and other settings are the same with MC-CDic-L. 

\begin{table}
	\centering
	\resizebox{8.6cm}{2.2cm}{
		\begin{tabular}{l|c|ccc|ccc}
			\toprule
			\toprule
			\multirow{2}{*}{Method} &\multirow{2}{*}{Params}& \multicolumn{3}{c}{IXI-T2}&\multicolumn{3}{c}{BrainTS-T2}\\
			\cline{3-8}
			&&PSNR$\uparrow$&SSIM$\uparrow$&RMSE$\downarrow$&PSNR$\uparrow$&SSIM$\uparrow$&RMSE$\downarrow$\\
			\midrule
			ZP&-&30.43&0.8840&8.15&32.40&0.8873&6.38 \\
			VDSR&0.7M&32.51&0.9162&6.37&35.53&0.9580&4.46 \\
			CSN&11.2M&34.04&0.9365&5.35&36.30&0.9633&4.08 \\
			RCAN&15.6M&34.44&0.9413&5.12&36.81&0.9670&3.84 \\
			SwinIR&11.9M&34.37&0.9402&5.17&36.99&0.9680&3.77 \\
			\midrule
			CUNet&0.2M&39.65&0.9720&2.77&38.17&0.9710&3.32 \\
			MCSR&3.5M&40.00&0.9734&2.67&38.09&0.9718&3.36 \\
			MASA&4.0M&41.45&0.9786&2.23&37.66&0.9700&3.50 \\
			MCMRSR&3.5M&40.93&0.9753&2.53&37.82&0.9660&3.45 \\
			MINet&11.9M&41.70&0.9784&2.22&39.46&0.9758&2.88 \\
			Ours-S&1.9M &\underline{41.96}&\underline{0.9792}&\underline{2.16}&\underline{39.87}&\underline{0.9773}&\underline{2.76} \\
			Ours-L&9.6M &\textbf{42.58}&\textbf{0.9810}&\textbf{2.01}&\textbf{40.75}&\textbf{0.9806}&\textbf{2.49} \\
			\bottomrule
			\bottomrule
		\end{tabular}
	}
\caption{Quantitative results of single-contrast MRI SR and multi-contrast MRI guided-SR algorithms when scaling factor is $\times4$.}
	\label{table-MCSR}
\end{table}
\subsection{Comparison with State-of-the-arts}
\begin{figure}
	\centering
	\includegraphics[width=1\columnwidth]{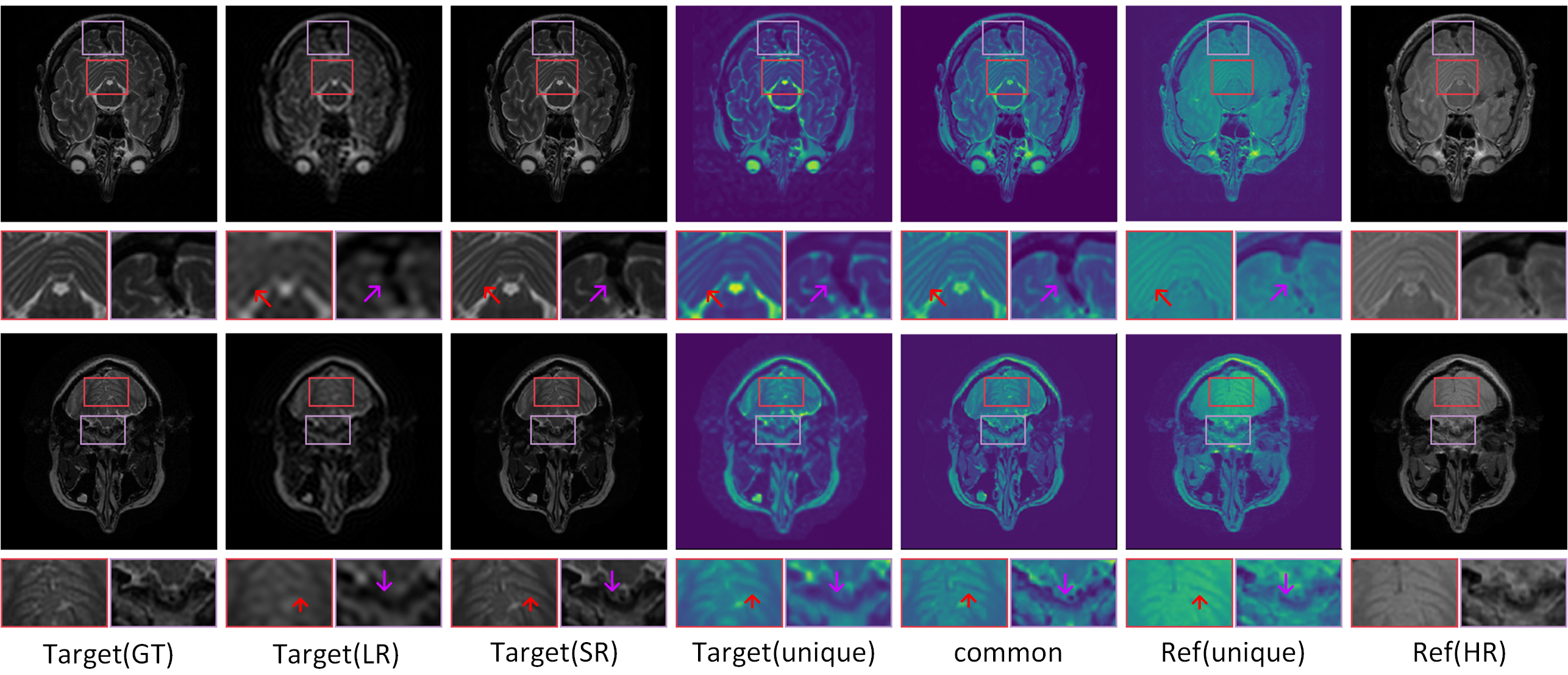} 
	\caption{Visualization of the decomposed common and unique components of our MC-CDic model on the guided-SR task.}
	\label{fig:decompose}
	\vspace{-2mm}
\end{figure}

\paragraph{Evaluations on Guided-SR.} Guided-SR aims to reconstruct the HR image from its LR image with the aid of another reference HR image. We compare our methods with various single image SR approaches, including VDSR~\cite{VDSR}, CSN~\cite{CSNet}, RCAN~\cite{RCAN} and SwinIR~\cite{swinir}, and various multi-contrast guided-SR methods, including MCSR~\cite{MCSR}, CUNet~\cite{CUNet}, MASA~\cite{MASA}, MINet~\cite{MINet} and MCMRSR~\cite{MCMRSR}. 
The quantitative results of these methods are shown in Table~\ref{table-MCSR}. Among these methods, our MC-CDic-L model gets the best performance. It outperforms MINet by 0.92dB and 1.29dB in PSNR on the IXI and BrainTS datasets, respectively. Our MC-CDic-S only has 1.9M parameters, but it outperforms most existing SOTA methods. We randomly select two images from the two test sets and visualize the error maps of different methods in Figure~\ref{fig:ref_sr}. As one can see, the results of our methods have fewer errors than other methods.
In Figure~\ref{fig:decompose}, we visualize the decomposed components of our MC-CDic model on the guided-SR task. The common feature represents the learned complementary information from the reference HR image to the target LR image. By decomposing the multi-contrast images, our method can only focus on the details (red arrows) related to the target image and avoid the interference of inconsistent information (purple arrows).

\begin{figure*}[t]
	\centering
	\includegraphics[width=1.8\columnwidth]{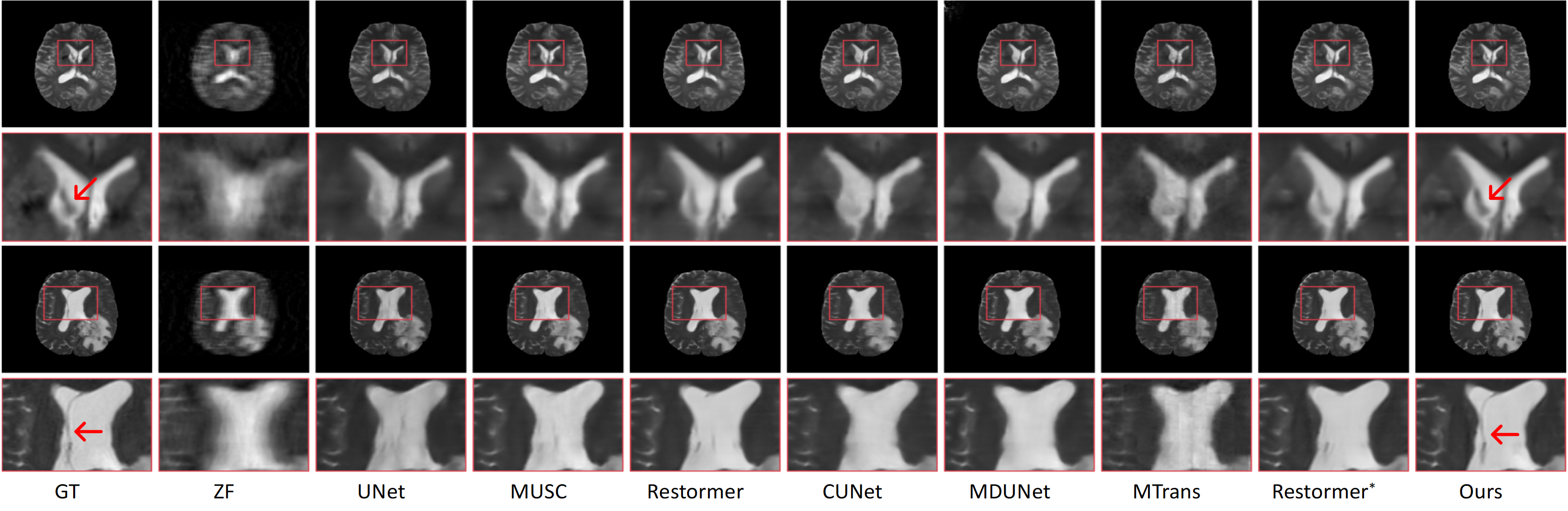} 
	\caption{Visual comparison of different single-contrast reconstruction (the first four methods) and multi-contrast guided reconstruction (the last five methods) on the IXI (PD guides T2) and BrainTS (T1 guides T2) testing sets when acceleration is $\times4$.}
	\label{fig:ref_rec}
\end{figure*}

\paragraph{Evaluations on Guided Reconstruction.}
Guided reconstruction aims to reconstruct a clear image from its under-sampled image with the aid of another auxiliary fully-sampled image. Here we compare our MC-CDic model with single-contrast reconstruction methods, including zero filling (ZF), UNet~\cite{unet}, MUSC~\cite{MUSC} and Restormer~\cite{restormer}, and multi-contrast guided-reconstruction methods, including MDUNet~\cite{xiang2018deep}, MTrans~\cite{CMMT} and Restormer$^*$~\cite{restormer}. Note that Restormer$^*$ employs multi-contrast images as input to restore the target contrast images.
Quantitative results are shown in Table~\ref{table-guided-rec}. From the table, we can find that our MC-CDic-L achieves the best performance on the two testing sets. Restormer$^*$ outperforms our MC-CDic-S, but it has far more parameters than our model. Visual comparisons are shown in Figure~\ref{fig:ref_rec}, our method can restore more anatomical details than other methods. It fully demonstrates the superiority of our method.

\begin{table}
	\centering
	\resizebox{8.6cm}{2cm}{
		\begin{tabular}{l|c|ccc|ccc}
			\toprule
			\toprule
			\multirow{2}{*}{Method} &\multirow{2}{*}{Params}& \multicolumn{3}{c}{IXI-T2}&\multicolumn{3}{c}{BrainTS-T2}\\
			\cline{3-8}&
			&PSNR$\uparrow$&SSIM$\uparrow$&RMSE$\downarrow$&PSNR$\uparrow$&SSIM$\uparrow$&RMSE$\downarrow$\\
			\midrule
			ZF&-&27.27&0.6188&11.69&28.72&0.6932&9.70 \\
			UNet&7.8M&33.24&0.9246&5.85&31.07&0.9443&7.83 \\
			MUSC&13.9M&34.39&0.9326&5.23&37.11&0.9644&3.71 \\
			Restormer&26.1M&35.79&0.9481&4.40&38.15&0.9699&3.31 \\
			\midrule
			CUNet&0.2M&37.30&0.9577&3.61&36.46&0.9629&4.04 \\
			MDUNet&5.2M&39.79&0.9715&2.75&35.26&0.9632&5.68 \\
			MTrans&86.1M&39.48&0.9700&2.96&34.02&0.9473&5.33 \\
			Restormer$^*$&26.1M&\underline{41.61}&\underline{0.9780}&\underline{2.24}&\underline{39.91}&\underline{0.9778}&\underline{2.75} \\
			Ours-S&1.9M&41.08&0.9765&2.37&39.75&0.9765&2.79 \\
			Ours-L&9.6M &\textbf{42.00}&\textbf{0.9802}&\textbf{2.14}&\textbf{41.65}&\textbf{0.9829}&\textbf{2.26} \\
			\bottomrule
			\bottomrule
		\end{tabular}
	}
\caption{Quantitative results of single-contrast MRI reconstruction and multi-contrast guided-reconstruction algorithms at $\times4$ acceleration.}
	\label{table-guided-rec}
\end{table}

\begin{figure}[t]
	\centering
	\includegraphics[width=1\columnwidth]{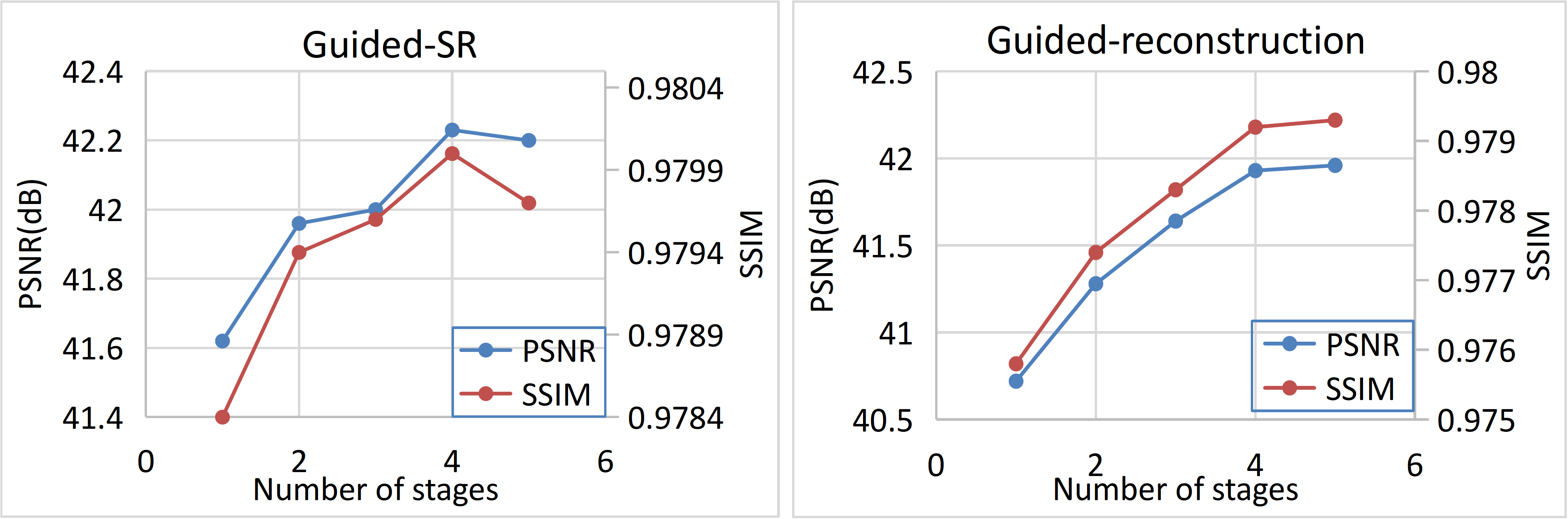} 
	\caption{The PSNR and SSIM curves on IXI validation set with a different number of stages $T$.}
	\label{fig:iter}
\end{figure}
\begin{figure}[t]
	\centering
	\includegraphics[width=1\columnwidth]{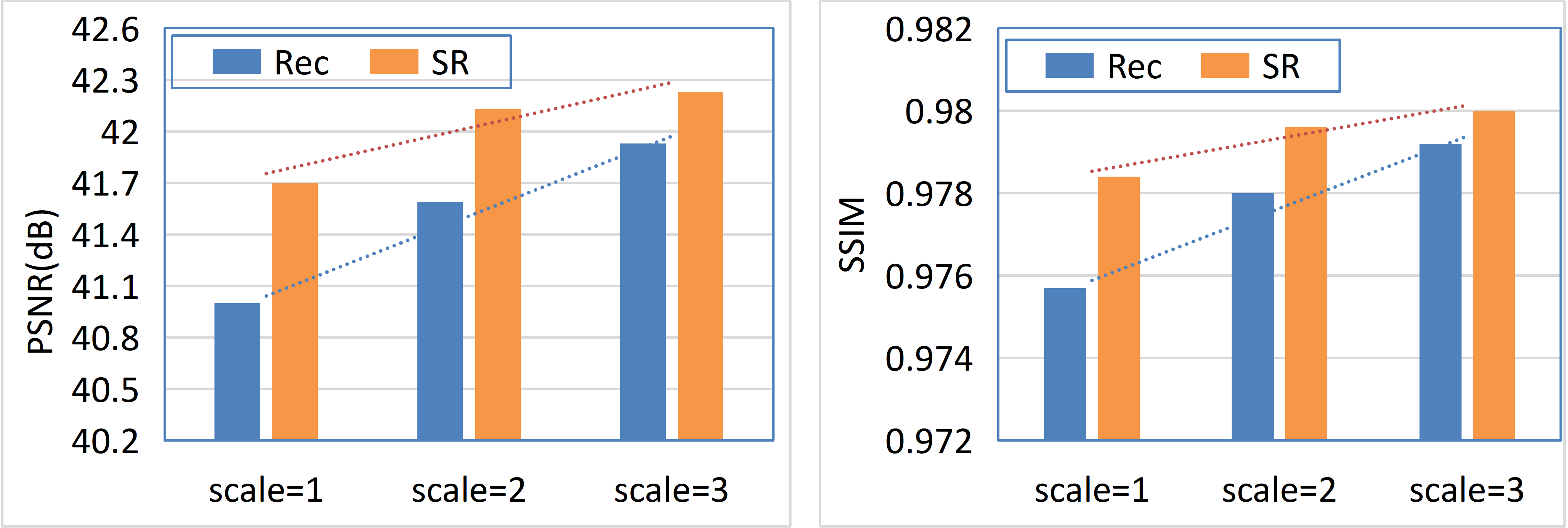} 
	\caption{PSNR and SSIM comparisons on IXI dataset with different levels of multi-scale dictionaries.}
	\label{fig:scale}
\end{figure}

\subsection{Ablation Analysis}
\paragraph{The Effect of the Number of Stages $T$.} To show how the number of iteration stages $T$ affects the reconstruction performance, we have compared the proposed MC-CDic method with different stages. Figure~\ref{fig:iter} shows the average PSNR and SSIM performance for the guided-SR task and guided-reconstruction task on the IXI validation set when $T\in[1,5]$.  It can be observed that more stages lead to improved reconstruction performance. Specifically, from $T=1$ to $T=4$, the psnr improves 0.6dB and 1.4dB in the guided-SR and guided-reconstruction tasks, respectively. To balance the performance and the computational complexity, we manually set $T=4$ in our final model.
\paragraph{The Effect of the Multi-Scale Dictionaries.} We compare the performance of the proposed MC-CDic model under different levels of multi-scale dictionaries in Figure~\ref{fig:scale}. From the figure, we can find that the performance increases with the feature levels.
From the trend lines, we can also see that multi-scale features are more important for the reconstruction task. Specifically, three levels of feature improve the PSNR by about 0.5dB and 0.9dB compared with the single-level features on the SR and reconstruction tasks, respectively. More feature levels may achieve better performance, however, the number of parameters and computations will also increase. To balance the performance and the computational complexity, we set the feature level as 1 in our MC-CDic-S and set the feature level as 3 in our MC-CDic-L. 
\paragraph{The Effect of the Proximal Network.} In this paper, we employ a proximal network to learn the prior information in a data-driven manner.
To validate its effectiveness, we compare the performance of the proposed MC-CDic model with and without the proximal network. Note that when we remove the proximal network, we use the soft-thresholding operator to constrain the learned features.
The performance of these two models is shown in Table~\ref{table-proxnet}. For the guided-SR task, the proximal network improves the performance by 0.43dB PSNR on the IXI-T2 dataset. For the guided-reconstruction task, the proximal network improves the performance by 1.06dB PSNR on the IXI-T2 dataset. It fully demonstrates the effectiveness of the proposed proximal network.

\begin{table}[t]
	\centering
	\resizebox{8cm}{0.9cm}{
		\begin{tabular}{c|ccc|ccc}
			\toprule
			\toprule
			\multirow{2}{*}{Method} & \multicolumn{3}{c}{Guided-SR}& \multicolumn{3}{c}{Guided-reconstruction}\\
			\cline{2-7}
			&PSNR$\uparrow$&SSIM$\uparrow$&RMSE$\downarrow$&PSNR$\uparrow$&SSIM$\uparrow$&RMSE$\downarrow$\\
			\midrule
			\emph{w/o} proxNet&42.15&0.9799&2.11&40.94&0.9770&2.42 \\
			\emph{w} proxNet&\textbf{42.58}&\textbf{0.9810}&\textbf{2.01}&\textbf{42.00}&\textbf{0.9802}&\textbf{2.14} \\
			\bottomrule
			\bottomrule
		\end{tabular}
	}
\caption{Quantitative results of the MC-CDic model with and without proximal network on IXI testing set.}
	\label{table-proxnet}
	\vspace{-2mm}
\end{table}

\section{Concluding Remarks}
We proposed a multi-scale convolutional dictionary model (MC-CDic) for multi-contrast MRI reconstruction and SR. 
Different from existing DCNN-based methods that manually designed fusion rules to fuse the multi-contrast information, our MC-CDic model was constructed under the guidance of the optimization algorithm with a well-designed data fidelity term. 
A proximal gradient algorithm was employed to solve the model and the iterative solutions were unfolded into a deep network. Especially, each module in the network could one-to-one correspond to an operation of a solution, thus our model was interpretable. We have tested our MC-CDic model on multi-contrast MRI reconstruction and SR tasks, experimental results demonstrate the superior performance of our method over the existing SOTA methods. In the future, we will explore more applications of our model on multi-modal tasks, such as RGB-guided depth map completion and RGB-guided depth map SR. 

\section*{Acknowledgements}
This work was supported by the National Key R\&D Program of China (2022ZD0161800), the NSFC-RGC (61961160734), the Shanghai Rising-Star Program (21QA1402500), and the Shanghai Municipal Commission of Economy and Informatization Program (2021-GZL-RGZN-01028).

\bibliographystyle{named}
\bibliography{ijcai23}
\end{document}